# Amorphous-silicon visible-light detector integrated in silicon nitride waveguides

Christian De Vita[1*], Fabio Toso[1], Natale Giovanni Pruiti[2], Charalambos Klitis[2], Giorgio Ferrari[1], Marc Sorel[2,3], Andrea Melloni[1], Francesco Morichetti[1]

[1]Department of electronics, information and bioengineering (DEIB), Politecnico di Milano, 20133 Italy

[2]University of Glasgow, Rankine Building, Oakfield Avenue, Glasgow G12 8LT, UK

[3]TeCIP Institute, Scuola Superiore Sant'Anna, 56124 Pisa, Italy

*Corresponding author: christian.devita@polimi.it

**Abstract** - Visible light integrated photonics is emerging as a promising technology for the realization of optical devices for applications in sensing, quantum information and communications, imaging and displays. Among the existing photonic platforms, high-index contrast silicon nitride ($Si_3N_4$) waveguides offer broadband transparency in the visible spectral range and a high scale of integration. As far as the complexity of photonic integrated circuits (PICs) increases, on-chip detectors are required to monitor their working point for reconfiguration and stabilization operations. In this work we present a compact in-line power monitor integrated in $Si_3N_4$ waveguides that operates in the red-light wavelength range (660 nm). The proposed device exploits the photoconductivity of a hydrogenated amorphous silicon (a-Si:H) film employed as a coating layer of the optical waveguide. Experimental results show a responsivity of 30 mA/W, a sensitivity of – 45 dBm and a sub-µs time response. These features enable the use of the proposed photoconductor for high-sensitivity monitoring and control of visible-light $Si_3N_4$ PICs.

Visible-light is a spectral region that is attracting increasing interest in many applications, including data communications [1], microscopy [2], biosensing [3], nanomedicine [4,5], quantum optics [6,7] and virtual reality [8]. In this wavelength range, silicon nitride ($Si_3N_4$) is one of the most established high-index-contrast platforms for photonic integrated circuits (PICs), because it offers a good tradeoff between low propagation loss waveguides and high integration scale [9]. As is happening in the near-infrared range, the route toward the realization of complex visible-light reconfigurable and programmable PICs [10] requires the integration of various photonic components on a single chip, including light sources and detectors. On-chip detectors are essential to monitor the working point of the PIC in order to set and stabilize its functionality by using closed-loop control tools. Different strategies have been recently proposed for the integration of visible-light photodetectors in $Si_3N_4$ waveguides. Monolithical integration of a p-i-n silicon photodiode at 488 nm has been demonstrated by exploiting vertical evanescent coupling between a $Si_3N_4$ waveguide and a crystalline silicon (c-Si) waveguide in a two-waveguide-layer silicon photonics platform [11]. Wide-band avalanche photodetectors at 685 nm have been monolithically integrated by end-fire coupling of a $Si_3N_4$ waveguide and a c-Si waveguide [12]. Although providing good performances respectively in terms of responsivity and speed, these approaches require complex technology steps to be performed on a pre-existing c-Si waveguide platform and cannot be ported on a conventional $Si_3N_4$ platform. Alternative options include the heterogeneous integration of Si or III-V compounds photodetectors, which require higher costs of fabrication and assembly.

Hydrogenated amorphous silicon (a-Si:H) is widely used in the visible range for thin-film light detectors, such as in photovoltaic applications, but in integrated optics it's use is rather limited. By introducing hydrogen during the a-Si deposition, dangling bonds are reduced and photoconductivity is significantly improved [13]. The integration of an a-Si detector on a lithium niobate on Si waveguide has been recently demonstrated at 850 nm wavelength [14]. In this work, we exploit the photoconductivity of an a-Si:H film deposited as a coating layer of a $Si_3N_4$ waveguide to realize a monolithically-integrated compact in-line photoconductor operating in the red-light range.

Figure 1(a) shows a schematic of the proposed device. The detector is integrated on a channel $Si_3N_4$ waveguide with a core thickness of 200 nm cladded with a 600-nm thick layer of hydrogen silsesquioxane (HSQ). The $Si_3N_4$ channel waveguide is fabricated on a 4 µm silica layer deposited by plasma enhanced chemical vapour deposition (PECVD) and is buried with a 600 nm thick hydrogen silsesquioxane (HSQ) upper cladding. The waveguide width $w$ = 400 nm guarantees single mode propagation down to a wavelength of 630 nm. The 200-nm $Si_3N_4$ film of the waveguide core was grown on a Si carrier wafer with 4 µm thermal oxide layer using Low Pressure Vapor Chemical Deposition (LPCVD). The waveguides were patterned by using electron beam lithography on a HSQ photoresist followed by reactive ion etching, and then they were coated with a 600-nm-thick HSQ layer cured for 72 h at 180 °C. The HSQ layer is locally lowered to a thickness $h_c$ in order to guarantee the proper overlap of the optical mode with the a-Si top-layer.

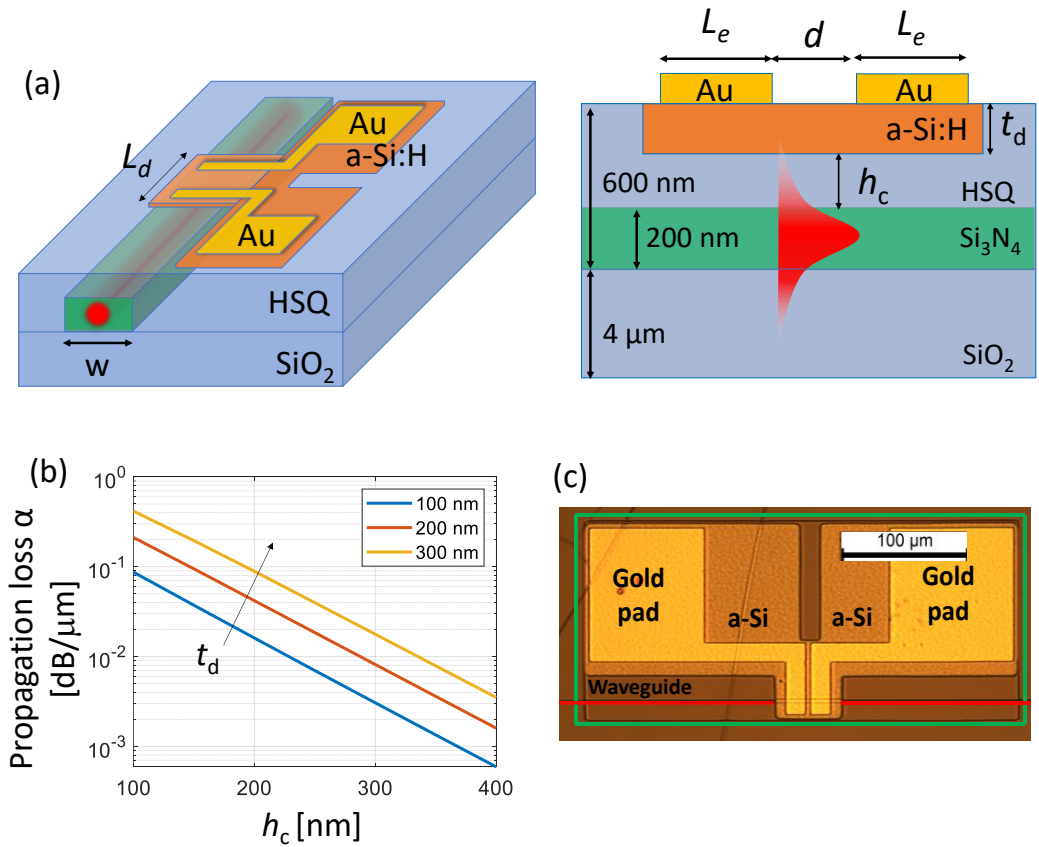

**Fig. 1.** (a) Schematic picture of the a-Si photoconductor integrated in a $Si_3N_4$ waveguide. (b) Numerical simulation of the propagation loss versus the spacing $h_c$ between a-Si:H film and the $Si_3N_4$ core of the waveguide for different thicknesses $t_d$ of the a-Si:H film. (c) Top-view microphotograph of one of the fabricated devices.

Numerical simulations based on the Beam Propagation Method (BPM) were carried out to evaluate the absorption of the a-Si layer versus the gap distance $h_c$. As refractive indices of $SiO_2$ and $Si_3N_4$ at a wavelength of 660 nm we used data available in the literature, $n_{Si3N4}$ = 2.007 [15], $n_{SiO2}$ = 1.471 [16]; the refractive indices and extinction coefficients of the a-Si and the cured HSQ were measured by using spectroscopic ellipsometry on test films, resulting respectively in $n_{HSQ}$ = 1.40, $n_{a-Si}$ = 3.60 and $k_{HSQ}$ = 0, $k_{a-Si}$ = 0.05. Figure 1(b) shows the simulated propagation loss α of the fundamental transverse electric (TE) guided mode versus $h_c$ for different a-Si thicknesses $t_d$. The choice of $h_c$ is a tradeoff between device footprint, insertion loss and responsivity of the detector. Using $t_d$ = 200 nm and $h_c$ = 200 nm (α = 0.04 dB/μm), an in-line detector with an overall length $L_d$ = 50 μm and 2 dB light absorption can be realized.

Figure 1(c) shows a top view microphotograph of one of the fabricated photodetectors. The HSQ cover is selectively lowered in the green rectangular area which is large enough (350 × 150 μm) to accommodate also the contact pads. The HSQ cover is selectively lowered in the green rectangular area which is large enough (350 × 150 μm) to accommodate also the contact pads. To this aim direct laser writing (DLW) lithography on a AZ5214E photoresist was used to protect the HSQ upper cladding outside the trench area from the successive RIE process (CHF$_3$ + O$_2$). Once the trench is opened, a 200 nm thick layer of a-Si:H was deposited by PECVD using SiH$_4$ as a gas precursor for suitable hydrogenation of the film [17], at a substrate temperature of 300 °C, chamber pressure of 500 mTorr and 50 W of RF power at 13.56 MHz. The a-Si:H film is then patterned to reduce the interaction length with the optical waveguide to 50 μm The metal contacts are made of 150-nm-thick gold evaporated on top of a 5-nm Cr adhesion interlayer and were patterned realized by lift-off., a width $L_e$ = 15 μm and are spaced by a distance $d$ = 4 μm (note that due to the limited resolution of the fabrication process, the length $L_d$ = 50 μm of the a-Si region is intentionally longer than the region covered by the electrodes, $2L_e + d$ = 34 μm).

A systematic analysis was carried out to estimate the propagation loss of the Si$_3$N$_4$ waveguide. All the results reported in this work refer to transverse-electric (TE) polarization. Figure 2 shows the transmission measurements on optical waveguides of increasing length at a wavelength of 660 nm. The employed test vehicles have a folded topology, as the one shown in the inset, where the length of the waveguide in the straight sections is varied, while the number and the radius of the bends are kept constant. The optical waveguides were edge-coupled with single-mode optical fibers (mode field diameter of 3.5 μm) without using mode adapters at the chip facet. For the 400-nm-wide waveguide employed for the detector, the fiber-to-waveguide coupling loss is about 10 dB/facet and the propagation loss is 1.88 dB/cm (+/- 0.1), while for a 350-nm-wide waveguide a slightly higher loss (2.06 dB/cm) is observed. These results are in line with state-of-the-art values reported in the literature [9].

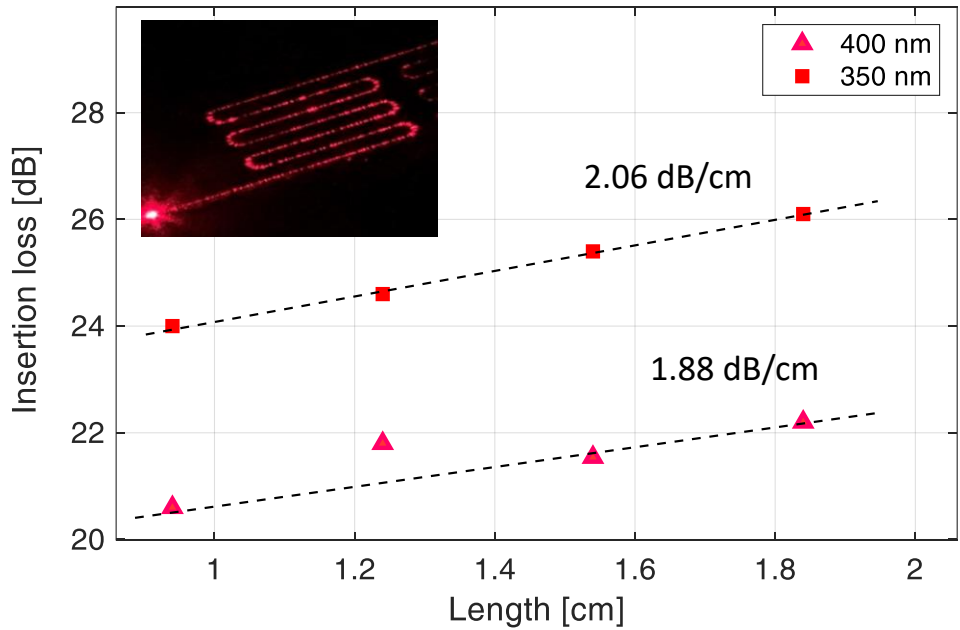

**Fig. 2.** Measured propagation losses at a wavelength of 660 nm of single mode Si$_3$N$_4$ waveguides width a width of 350 nm (red squares) and 400 nm (pink triangles). Inset: photo of the chip

For the assessment of the photoconductor response, the photonic chip was wire-bonded onto a thermally-stabilized printed circuit board (PCB). Figure 3(a) shows the light-dependent current measured at the two electrodes for increasing values of the optical power $P_d$ at the input of the detector, when a constant voltage $V_e$ is applied at the two electrodes. The reference dark current is about 25 pA at a $V_e$ = 4 V and 50 pA at $V_e$ = 8 V, corresponding to a resistance of 160 G$\Omega$ between the electrodes. Since the transversal width of the a-Si film between the electrodes is 70 µm, this results in a conductivity of 18 nS/cm for the a-Si:H film. The dark current limits the sensitivity of the device to a minimum detectable power of -45 dBm. Above the sensitivity threshold, we observe a change of the photocurrent versus the optical power $P_d$ across a dynamic range of more than 20 dB, the maximum power of -25 dBm being limited by the loss of the experimental setup. The responsivity $R_{ph}$ of the device, shown in Fig. 3(b) is defined with respect to the absorbed power $P_{abs} = \eta P_d$, where $\eta = 1-\exp(-\alpha L_d)$ is the fraction of the guided light absorbed by the a-Si film. Loss measurements performed on a-Si coated test waveguides show that a 50-µm-long section introduces a loss of about 2 dB, in agreement with the numerical results of Fig. 1(c). This means that the detector absorbs 37% of the guided light. Results in Fig. 3(a) show that, when $P_d$ = –30 dBm, corresponding to $P_{abs}$ = 370 nW, a photocurrent of about 11 nA is measured for $V_e$ = 8V, resulting in a responsivity $R_{ph}$ of about 30 mA/W. This photocurrent essentially flows in the a-Si region on top of the waveguide that overlaps with the guided mode. Assuming a transversal width equal to that of the waveguide core (400 nm), we estimate an increase of the a-Si conductivity to 0.68 mS/cm, that is more than 4 orders of magnitude with respect to the dark level value (18 nS/cm).

Some considerations need to be made on the device responsivity $R_{ph}$. First, in Fig 3(b) we see that $R_{ph}$ remains almost constant versus the optical power $P_d$, meaning that the photocurrent increases linearly with $P_d$. Actually [see Appendix 2 for a linear scale representation of Fig. 3(a)]. This effect implies that in the considered optical power range (from -45 to -25 dBm) neither saturation effects nor carrier lifetime reduction occur, which can arise in the a-Si film at high photon flux through [18]. Second, in a photoconductor the photocurrent is expected to increase linearly with the applied voltage $V_e$, but in the measured device when $V_e$ is halved from 8 V to 4 V, the responsivity $R_{ph}$ reduces by three times from about 30 mA/W to 10 mA/W. However, this behavior is consistent with the I-V curves of Fig. 2(b) indicating the presence of a Schottky diode at the a-Si-metal interface, with a threshold voltage $V_{th}$ = 2 V. The presence of non-ohmic contacts implies that the effective voltage applied to the a-Si film in the detector region is actually $V_{eff} = V_e - V_{th}$, that is 6 V and 2 V instead of 8 V and 4 V, respectively.

Regarding the carrier lifetime, in amorphous semiconductors absorption of photons with an energy higher than the material band gap produces different kind of carriers that can contribute to the overall photocurrent. These include free carriers in the conduction (electrons) and valence (holes) bands as well as trapped carriers associated with intragap defects states or surface states. In a-Si trapped carriers' lifetime may be orders of magnitude higher than free carrier lifetime [18].

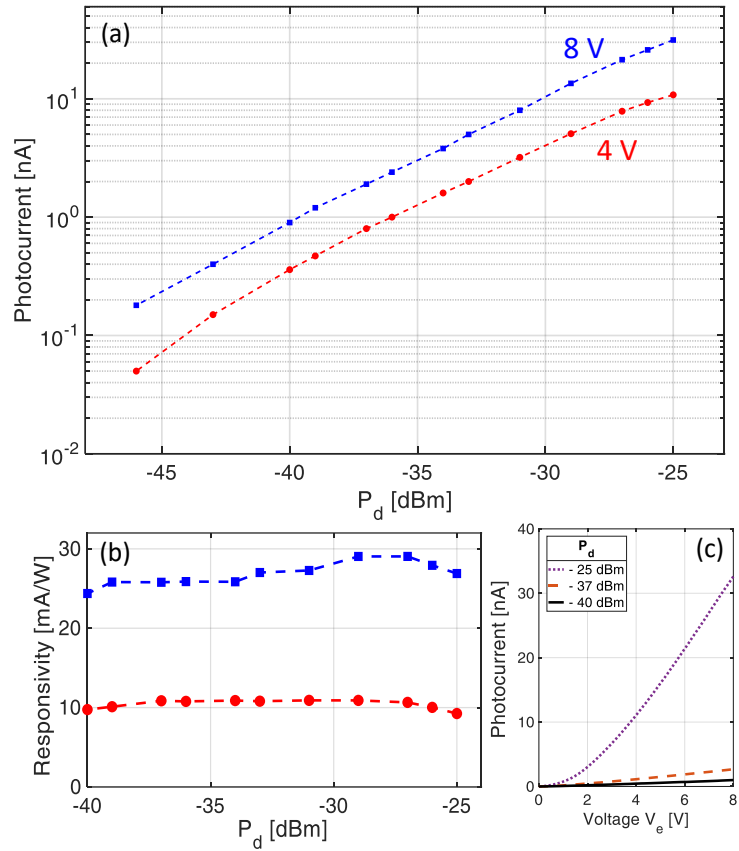

**Fig. 3.** (a) Measured photocurrent for increasing optical power $P_d$ at a wavelength of 660 nm when the applied voltage is $V_e$ = 4 V (red squares) and 8 V (blue circles). (b) Responsivity from measurements shown in (a). (c) IV curve for increasing power $P_d$.

An effective lifetime $\tau_{eff}$, which depends on the electric field profile between the electrodes and the density of defect states [19] can be defined. The lifetime-mobility product $\mu\tau_{eff}$ is a characteristic parameter to qualify the properties of a photoconductor and can be directly derived from the responsivity $R_{ph}$ as (see derivation in Appendix 1)

$$\mu\tau_{eff} = R_{ph} \frac{h\nu}{q} \frac{L_d d}{V_{eff}} \qquad (1)$$

where $q$ is the electron charge, $h$ is the Planck constant and for simplicity we assumed a uniform electric field $V_{eff}/d$ in the gap region between the electrodes. In our device $\mu\tau_{eff}$ is in the order of 2e-8 [cm$^2$/V] (see Appendix 2, fig. AP2), which in line with data reported in previous works for a-Si films [20]. The photoconductive gain $G_{ph}$ is given by

$$G_{ph} = \frac{R_{ph}h\nu}{q} = \mu\tau_{eff}\frac{V_{eff}}{L_d d} = \frac{\tau_{eff}}{\tau_t}\frac{d}{L_d} \qquad (2)$$

where $\tau_t = d^2/\mu V_{eff}$ is the transit time of the carriers though the a-Si:H film. A rather low photoconductive gain $G_{ph}$ of about 0.05 at $V_e$ = 8 V and 0.02 at 4 V (see Appendix 2, Fig. AP3) is observed which is limited by both the geometrical ratio $d/L_d$ = 0.08 and the time constants ratio $\tau_{eff}/\tau_t$ = 0.69 at $V_{eff}$ = 6 V. To quantify the magnitude of the two-time constants, we can assume a mobility $\mu$ in the order of 1 cm$^2$/V·s for the a-Si film [21], so that we obtain $\tau_{eff}$ = 18 ns and $\tau_t$ = 26 ns ($V_{eff}$ = 6 V, $P_d$ = –30 dBm).

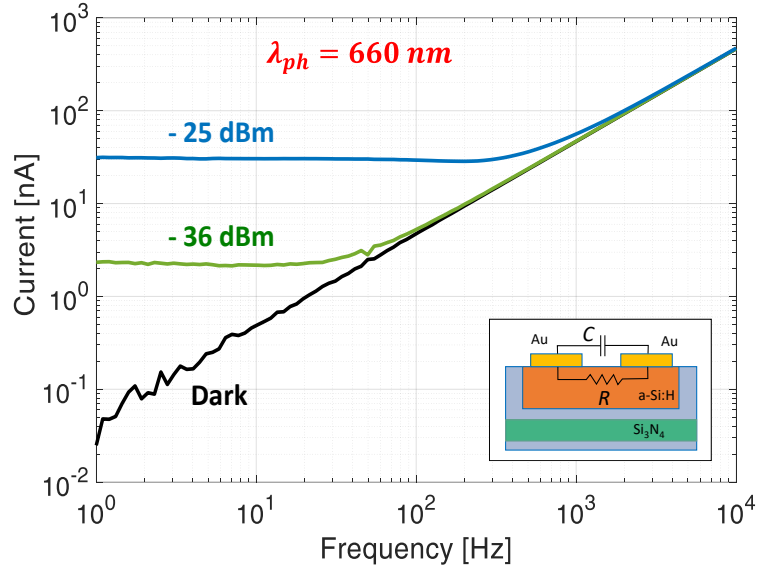

**Fig. 4.** Measured photocurrent vs frequency for different levels of the optical power $P_d$ at the detector. $V_e$= 8 V, $V_{eff}$=6 V

We expect that the photoconductive gain $G_{ph}$ can be increased by more than one order of magnitude by reducing the electrode width to $L_e \sim 1$ µm as well as by reducing the transit time $\tau_t$ if a shorter gap $d$ between the electrodes is realized (see Appendix 2 for more considerations about responsivity, gain and mobility-lifetime product).

To investigate the frequency-domain response of the photoconductor, we measured the change of the electrical impedance at the electrical pads of the device versus the light power $P_d$. The impedance measurement was performed with a lock-in detection scheme, where a sinusoidal voltage with peak value $V_e$ = 8 V at a frequency $f_c$ was applied to one electrode, while the other electrode was connected to ground. The electrical equivalent circuit of the detector is shown in the inset of Fig. 4 and consists of an $RC$ circuit, where $R$ is given by the a-Si film between the electrodes, whereas $C$ is the parasitic capacitance between the electrodes (including also the bonding wires and pads). The lock-in amplifier employed for the characterization in frequency has a minimum noise floor of 20 fA at 1 Hz integration bandwidth.

Figure 4 shows the measured AC current versus frequency $f_c$ for increasing values of $P_d$. The plateau in the low-frequency side of the curves indicates the amplitude of the photo current flowing through the light-dependent resistance $R$ of the a-Si:H film. In agreement with the results of Fig. 3, the photocurrent is about 30 nA for an optical power of -25 dBm while the reference dark current (black line) is less than 100 pA. At higher frequencies all the curves asymptotically approach the straight line provided by the current flowing through the parasitic capacitance $C$, that is in the considered device is about 922 fF. This effect limits the visibility of the frequency domain response of the detector to 1 kHz at -25 dBm.

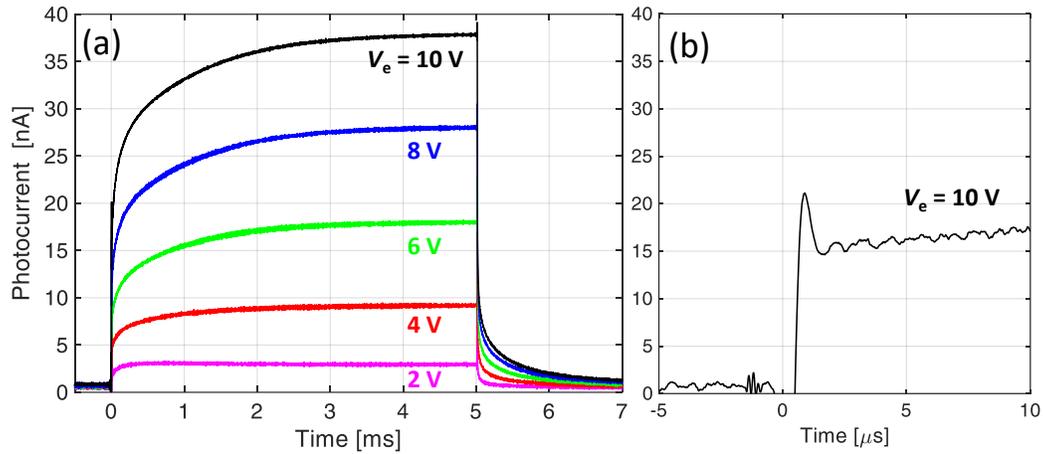

**Fig. 5.** a) Time-domain response of the integrated a-Si photodetector for different bias voltage $V_e$. A fast response due to free-carrier dynamics (sub-µs time scale) is followed by a slower dynamics (ms-scale) due to intra-gap trapped carriers. b) sub-µs time response for $V_e$= 10 V.

The time-domain response of the a-Si photoconductor was assessed by using the device to detect an intensity modulated optical signal propagating in the $Si_3N_4$ waveguide. The signal is generated by directly modulating a 660-nm visible light laser source with a square wave driving current with a pulse repetition rate of 100 Hz. The dynamics of the laser was characterized with a fast silicon photodiode with a rising time of 9 ns, showing that the rise time of the laser is about 1 µs. The read out of the a-Si photoconductor was performed with a low noise transimpedance amplifier providing a gain of $10^8$ A/V. The output of the TIA was collected by an oscilloscope triggered by the electrical signal modulating the laser source. Figure 5 shows the time response of the a-Si photoconductor when $P_d$ = - 25 dBm for increasing voltage $V_e$. Two different time constants are visible in the dynamics of the device. A slower time constant, in the range of several hundreds of µs, is due to the carriers trapped in the intragap states of a-Si [22]. However, as shown in the inset of the figure, a very fast response is observed in the first part of the transient, which is associated to the effective lifetime of free carriers $\tau_{eff}$ and is expected to occur in a time scale of several ns. In our experiment the observation of the fast time response is limited by the rise-time response of the laser (~1 µs) and by the limited bandwidth of the TIA that is responsible for the damping of the measured current signal.

In conclusion, we reported on an a-Si in-line photoconductor operating in the 660-nm wavelength range that is monolithically integrated with standard $Si_3N_4$ waveguides. The device has a sensitivity of -45 dBm, a responsivity of 30 mA/W and a good linearity across a dynamic range of more than 20 dB (limited by the experimental setup). The time- response of the device is limited to a few ms time scale by the dynamics of intra-gap trapped carriers in the a-Si film, but a pronounced sub-µs response is observed that can be exploited for fast monitoring operations for the control and calibration of visible-light $Si_3N_4$ PICs. The size of the device (50 µm) could be downscaled by at least one order of magnitude by using higher-resolution lithography, thus leading also to a 10x improvement of the responsivity. Due to the well-assessed photogeneration properties of a-Si, which extend across the entire visible light range, the presented photoconductor concept can be extended to lower wavelengths to implement RGB detectors integrated in $Si_3N_4$ waveguides. Finally, the additive CMOS-compatible fabrication process makes the device concept portable to any photonic platform for visible light applications.

**Appendix 1**

**Analytical derivation of µτ$_{eff}$ and the responsivity R$_{ph}$**

Material conductivity is generally expressed as

$$\sigma = q\,\mu\,N$$

where $q$ is the elementary charge, $\mu$ is the carrier mobility and $N$ is the carrier density. In stationary conditions the photogenerated carriers are given by

$$N = G\,\tau_{eff},$$

where $G$ is the generation rate and $\tau_{eff}$ is the effective recombination time.

Considering a variation in the conductivity $\Delta\sigma$, due to the incident light, we can write that

$$\Delta\sigma = q\,\mu\,\Delta N = q\,\mu\,\Delta G\,\tau_{eff},$$

Assuming unitary internal quantum efficiency, $\Delta G$ depends on the number of absorbed photons per volume

$$\Delta G = \frac{P_{abs}}{h\nu} \cdot \frac{1}{A\,L_d},$$

where $P_{abs}$ is the optical absorbed power, A is the transverse cross-section and $L_d$ is the length of the amorphous silicon region interacting with the waveguide mode.

Now, from the first Ohm's law we can write that

$$V_{eff} = RI = \frac{1}{\sigma}\frac{d}{A}I = \frac{h\nu\,A\,L_d}{P_{abs}\,q\,\mu\,\tau_{eff}}\frac{d}{A}I$$

From which we can write the mobility-lifetime $\mu\tau_{eff}$ product as function of the responsivity $R_{ph} = I/P_{abs}$ and the dimensions of the detector

$$\mu\tau_{eff} = \frac{h\nu\,L_d\,d}{V_{eff}P_{abs}\,q} \cdot I = \frac{h\nu\,L_d\,d}{V_{eff}q} \cdot R_{ph} = R_{ph}\cdot\frac{h\nu}{q}\cdot\frac{L_d d}{V_{eff}}.$$

**Appendix 2**

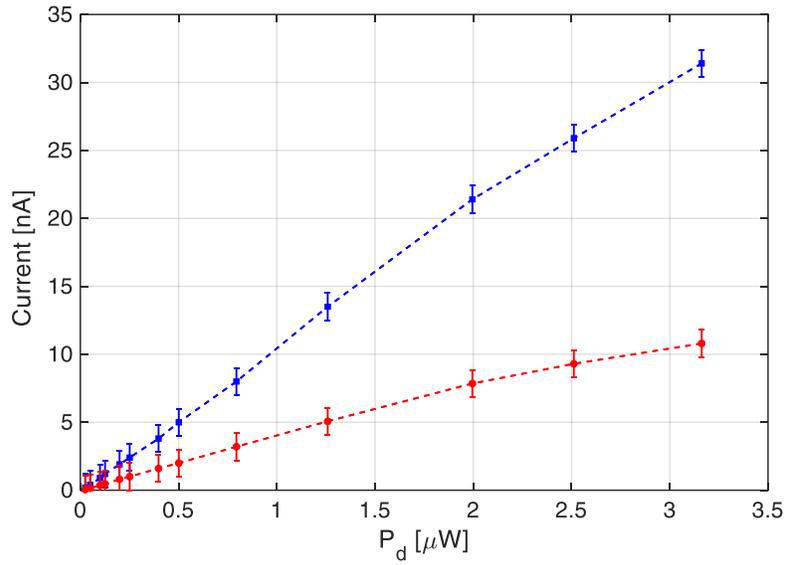

Figure AP2. Measured photocurrent for increasing optical power $P_d$ at wavelength of 660 nm when the applied voltage is $V_e$ = 4 V (red squares) and 8 V (blue circles). An error bar of ±1 nA is added representing the sensitivity of the used instrumentation.

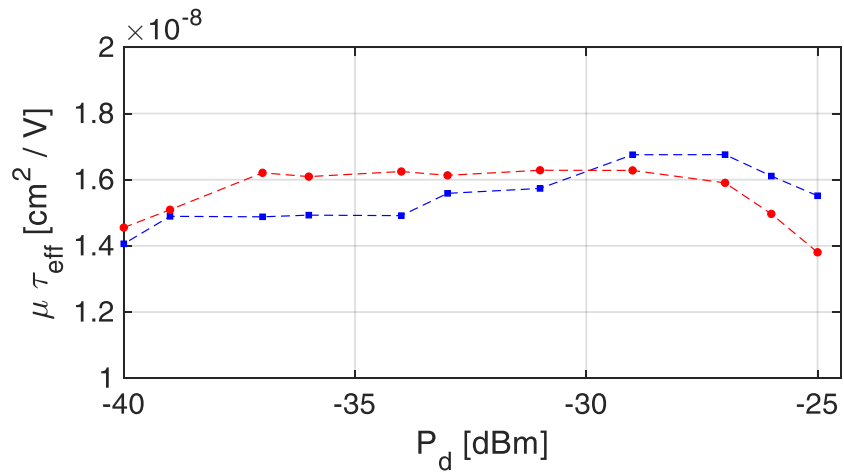

Figure AP2. Mobility-lifetime product extracted from from data in Fig. 1

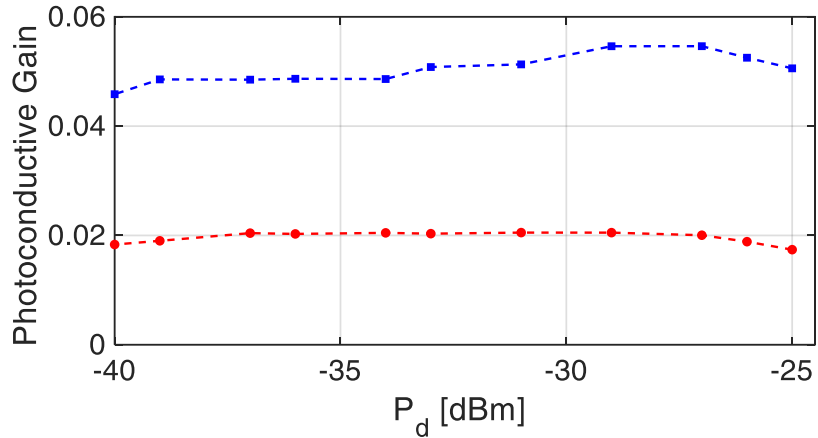

Figure AP3. Photoconductive gain extracted from data in Fig. 1.